\documentclass[prl,twocolumn]{revtex4}%
\usepackage{amsmath}
\usepackage{amsfonts}
\usepackage{amssymb}
\usepackage{graphicx}%
\begin{document}
\title{Low temperature solution of the Sherrington-Kirkpatrick model}
\author{Sergey Pankov}
\affiliation{National High Magnetic Field Laboratory, Florida
State University, Tallahassee, FL 32306}
\date{\today{}}


\begin{abstract}
We propose a simple scaling ansatz for the full replica symmetry
breaking solution of the Sherrington-Kirkpatrick model in the low
energy sector. This solution is shown to become exact in the 
limit $x\to0$, $\beta x\to\infty$ of the Parisi replica symmetry 
breaking scheme parameter $x$. The distribution function $P(x,y)$ 
of the frozen fields $y$ has been known to develop a linear gap 
at zero temperature. We integrate the scaling equations to find 
an exact numerical value for the slope of the gap 
$\partial P(x,y)/\partial y|_{y\to0}=0.3014046...$ We also 
use the scaling solution to devise an inexpensive numerical 
procedure for computing finite timescale ($x=1$) quantities. 
The entropy, the zero field cooled susceptibility and the local 
field distribution function are computed in the low temperature 
limit with high precision, barely achievable by currently 
available methods.

\end{abstract}
\maketitle

The field of spin glass physics has been actively studied for the
last thirty years. Much attention has been devoted to the
Sherrington-Kirkpatrick (SK) model \cite{sherrington75}--a mean
field analog of realistic spin glasses. The mean field treatment
of this model, although exact, is highly
nontrivial \cite{parisi79}. In its static formulation, the mean
field description of the glassy phase involves infinitely
many steps of replica symmetry breaking (RSB), 
compactly written as a set of integro-differential
equations \cite{parisi80} for the spin glass order parameter $q(x)$
and some auxillary functions \cite{almeida83} $m(x,y)$ and $P(x,y)$.
A dynamical formulation \cite{sompolinsky&zippelius81} leads to
similar results and interprets the variable $x\in[0,1]$ as a
parametrization of (diverging) timescales of the model, with
smaller values of $x$ corresponding to the longer
times \cite{sompolinsky81}. The functions $m(x,y)$ and $P(x,y)$ are
interpreted \cite{sommers84}, respectively, as a local
magnetization in the presence of a frozen field $y$ and a
distribution function of the frozen fields, measured at a
timescale $x$. The full RSB equations cannot be solved analytically, 
and even solving them numerically is not a trivial task. 
There were attempts to conjecture certain scaling
laws \cite{parisi&toulouse80}, which, if correct, would grant an
exact solution of the problem. Those conjectures, however, turned
out to be merely a good approximation \cite{crisanti03}.

The idea of scaling, nevertheless, deserves attention. The spin glass
phase is characterized by marginal criticality, and critical
systems generically exhibit some kind of universal (scaling)
behavior \cite{pazmandi99}. To be more specific we use insights
gained from both the static and dynamic viewpoints of the problem.
The space of spin glass states is ultrametric and can be
visualized as a tree of states \cite{mezard84} with the leaves
representing quasi-equilibrium (pure) states. In the dynamical
description the system can explore the tree, reaching the states
with smaller and smaller overlap $q$, connected to the initial
state through larger and larger branches. The self-similarity of
the tree, away from its leaves ($x\sim1$) and its root 
($x\lesssim T$), is an important ingredient which could be related 
to the scaling behavior of the functions $m(x,y)$ and $P(x,y)$ in 
the range $T\ll x\ll1$. Another condition for universality to occur 
is that the typical frozen fields should play no role in the problem.
In other words, the scale of disorder should be irrelevant. This 
can only happen in the zero temperature limit, when any finite 
frozen field completely polarizes a spin, thus excluding self from 
contributing to the dynamical evolution of the system. Therefore, 
it is the limit of $x\to0$ and $\beta x\to\infty$, where one should 
expect to find a scaling solution, which would be valid for $y\ll1$. 
In this paper we present such a solution and prove that it becomes
exact in the above limit, by investigating the correction to scaling. 
We then use the scaling solution to solve the Parisi equations on 
the scale $y\sim T$ and $x>0$, for the first time directly in the 
zero temperature limit. Various physical quantities, measurable on
short time scales, are also computed.

\emph{Model.} We consider the Sherrington-Kirkpatrick 
model--perhaps the most studied model of spin glass physics:
\begin{equation}
H=\sum_{i<j} J_{ij}s_is_j,
\label{skmodel}
\end{equation}
where the spins are classical variables $s_i=\pm1$ and the bonds
are randomly Gaussian quenched with a variance $J$, which is set 
to unity through the rest of the paper. Upon cooling, the ergodicity 
breaks down at $T=1$ and the systems freezes in one of the multitude 
of quasiequilibrium states. This phase is referred to as the glassy 
phase and can be described with use of the Parisi RSB formalism.
A spin on a site $i$ experiences a local field $\sum_{j}J_{ij}s_j$.
The local fields can be measured with or without the site $i$ 
present in the lattice. They are called, accordingly, 
the instantaneous or frozen fields. 
For $T<1$ a pseudogap in the 
distribution of fields opens, becoming (asymptotically) linear
in the limit $y\to0$, $\beta y\to\infty$.

\emph{Scaling ansatz.} The Parisi infinite RSB scheme equations,
written in a usual form \cite{sommers84,thomsen86}, read:
\begin{eqnarray}
\dot m(x,y)=-\frac{\dot q(x)}{2}\left[m''(x,y)+2\beta x
m(x,y)m'(x,y)\right],
\label{diffeqm}\\
\dot P(x,y)=\frac{\dot q(x)}{2}\left\{P''(x,y)-2\beta x
\left[m(x,y)P(x,y)\right]'\right\}, \label{diffeqP}
\end{eqnarray}
where the dot and prime are derivatives with respect to $x$ and
$y$ variables, correspondingly. The differential equations are
supplemented with the initial conditions: $m(1,y)=\tanh{\beta y}$
and $P(0,y)=\delta(y)$. In principle, these equations can be
solved iteratively. One can compute $m$ and $P$ for a given order
parameter $q$, which, in its turn, is computed from $m$ and $P$:
\begin{equation}
q(x)=\int dy P(x,y)m^2(x,y).
\label{inteqq}
\end{equation}
We introduce new notations: $m(x,y)=\tilde m(x,z)$, $P(x,y)=(\beta
x)^{-1}\tilde p(x,z)$, $\dot q(x)=2\beta (\beta x)^{-3} c(x)$,
where $z=\beta xy$. It will become clear later that in the scaling
regime the functions $\tilde m$, $\tilde p$ and $c$ lose their
dependence on the variable $x$. We recast 
Eqs.~(\ref{diffeqm},\ref{diffeqP}) using new definitions:
\begin{eqnarray}
& x\dot{\tilde m}=-c\left(\tilde m''+2\tilde m\tilde m'\right)-z
\tilde m',
\label{diffeqmt}\\
& x\dot{\tilde p}=c\left[\tilde p''-2(\tilde p\tilde m)'\right]
-z\tilde p'+\tilde p,
\label{diffeqpt}
\end{eqnarray}
where the dot and prime are now derivatives with respect to $x$
and $z$ variables, correspondingly. 
The functions' arguments are omitted for compactness.
Our scaling ansatz states that $\tilde m$ and $\tilde p$ are 
functions of the scaling variable $z$ only, and $c$ is a constant, 
given by $c=\int dz\tilde p(z)(1-\tilde m^2(z))$, 
as follows from Eq.~(\ref{inteqq}). Hereafter we refer to 
Eqs.~(\ref{diffeqmt},\ref{diffeqpt}), with the left hand sides 
set to zero and $c$ being a constant, as the scaling equations.

\begin{figure}[ptb]
\includegraphics[width=3.3in]{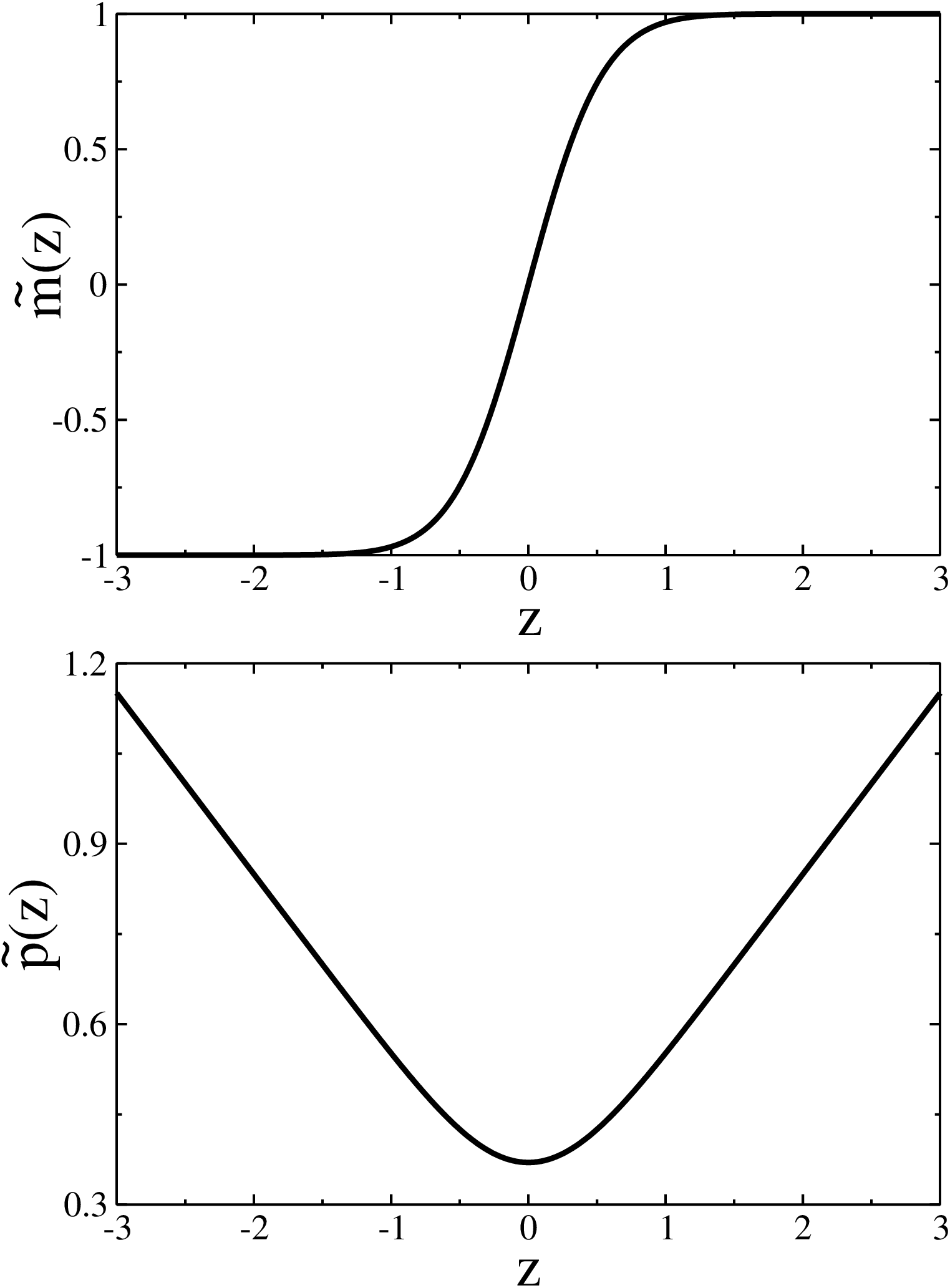} 
\vspace{.1cm}
\caption{Scaling functions $\tilde m(z)$ and $\tilde p(z)$, 
obtained by integrating the Parisi equations 
[Eqs~(\ref{diffeqmt},\ref{diffeqpt})] under the assumptions of the 
scaling ansatz. The field distribution function is asymptotically 
linear, $\tilde p(z)=\gamma|z|+2\gamma c$, $|z|\gg1$.}
\label{mtptfig}
\end{figure}

As expected, the ansatz does not respect the boundary conditions
(where the tree of states is not self-similar),
but we will demonstrate that it becomes asymptotically exact in
the scaling regime described above. This is precisely where the
slopes of the linear gap are formed. Thus the scaling ansatz
allows us to compute the slope of $P(1,y)$ in the low temperature
limit. Because the scaling equations can be easily integrated
(numerically), one can obtain the value of the slope with 
arbitrary precision. The initial conditions for the scaling 
equations follow from the definition of $m$ and $P$ and from
the linearity of the gap at large $z$:
\begin{align}
& z=0,\quad \tilde m=0, \quad \tilde p'=0;
\label{initcondz0}\\
& |z|\gg1,\quad \tilde m={\rm sign}(z), \quad \tilde
p=\gamma(|z|+2c).
\label{initcondzinf}
\end{align}
With a fixed constant $c$ the function $\tilde p$ enters the
equations linearly, therefore one can set $\gamma=1$ when solving
for $c$, and then compute the slope as $\gamma=c [\int dz \tilde
p_{\gamma=1}(z)(1-\tilde m^2(z))]^{-1}$. Up to ten digits of
precision we found $c=0.4108020997$, $\gamma=0.3010464715$. The
functions $\tilde m(z)$ and $\tilde p(z)$ are shown in 
Fig.~(\ref{mtptfig}).

\begin{figure}[ptb]
\includegraphics[width=3.3in]{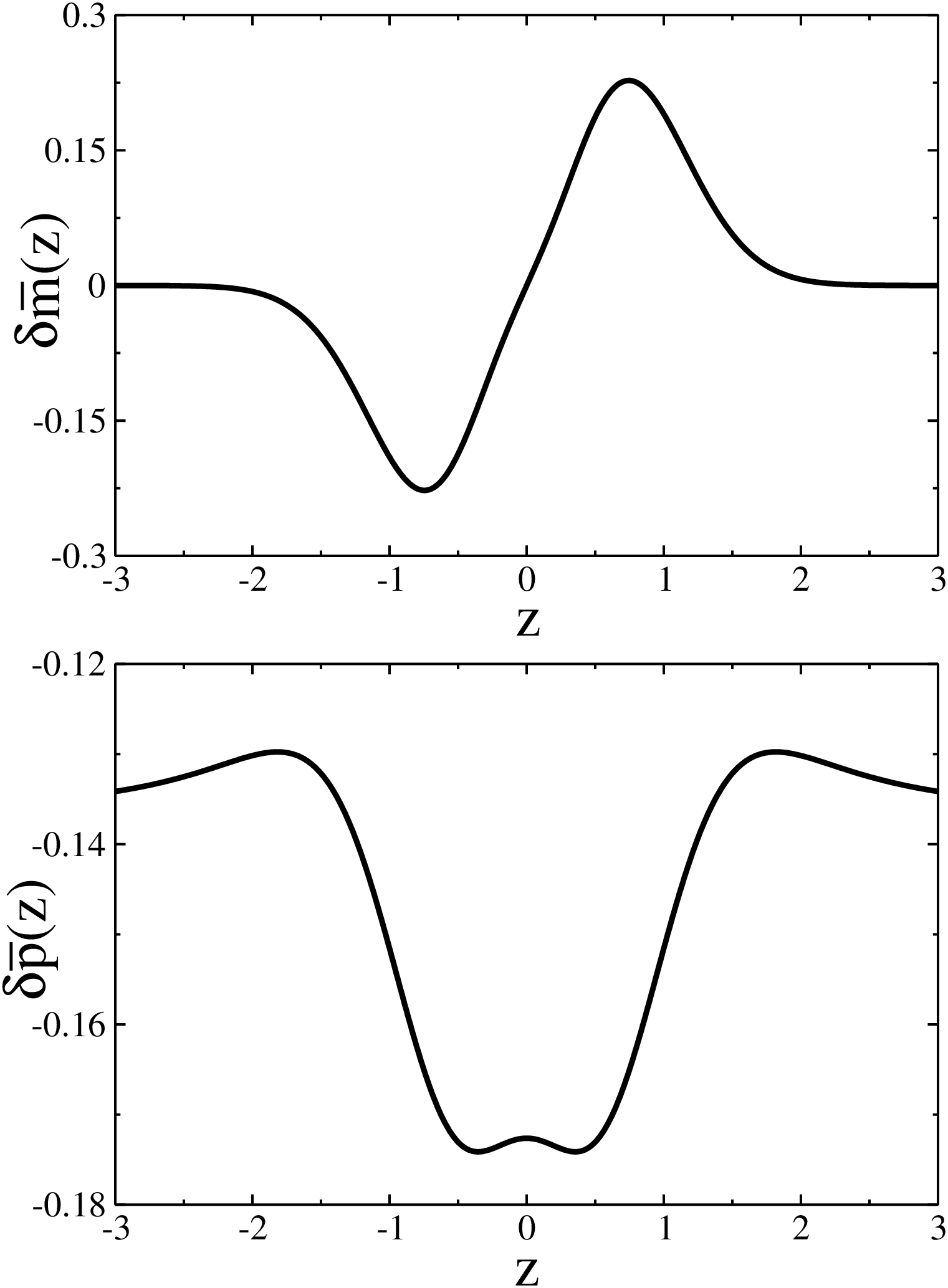} 
\vspace{.1cm}
\caption{Normalized correction to scaling functions 
$\delta\bar m(z)=\delta\tilde m(x,z)/\delta c(x)$ and
$\delta\bar p(z)=\delta\tilde p(x,z)/\delta c(x)$, obtained
from the linear stability analysis 
Eqs~(\ref{lindiffeqm}-\ref{lininteq1}). Correction to scaling
terms exhibit quick decay with $x$, as $\sim x^{5.41}$ - 
a manifestation of good quality of the Parisi-Toulouse
scaling approximation.}
\label{dmdpfig}
\end{figure}

\emph{Stability of the scaling solution.} To substantiate our
findings we have to prove the existence of the scaling regime. 
Equations~(\ref{diffeqmt},\ref{diffeqpt}) are free of
singularities, and because the scaling solution differs from the
initial conditions at $x=1$, our scaling [coinciding with
Parisi-Toulose scaling \cite{parisi&toulouse80} for $q(x)$] cannot
hold for all $x$. The important question is if there is at all
such an $x$ where the proposed scaling solution becomes exact. One
can answer this question by investigating the stability of the
scaling solution, expanding around it to linear order. 
We look for a correction to scaling in the form: 
$\delta\tilde m(x,z)=x^{\lambda}\delta\tilde{\tilde
m}(z)$, $\delta\tilde p(x,z)=x^{\lambda}\delta\tilde{\tilde m}(z)$
and $\delta c(x)=x^{\lambda}\delta{\tilde c}$. This procedure
leads to a set of inhomogeneous linear differential equations
\begin{align}
& \lambda\delta\bar m=-c\left[\delta\bar m''+2(\delta\bar m \tilde
m)'\right]-z\delta\bar m'+z\tilde m'/c,
\label{lindiffeqm}\\
& (\lambda-1)\delta\bar p=c\left[\delta\bar p''-2(\delta\bar p
\tilde m)'-2(\tilde p \delta\bar m)'\right] \nonumber\\
&\qquad\qquad\qquad\qquad\qquad -z\delta\bar p'+(z\tilde p'-\tilde
p)/c,
\label{lindiffeqp}
\end{align}
supplemented by two constraints:
\begin{align}
 & 1=\left(1-\frac{\lambda}{2}\right)\int dz
\left[\delta\bar p(1-\tilde m^2)-2\tilde p\tilde m\delta\bar
m\right],
\label{lininteqq}\\
& 0=\int dz \left[\delta\bar p(\tilde m')^2+2\tilde p\tilde
m'\delta\bar m'\right], \label{lininteq1}
\end{align}
where $\delta\bar m=\delta\tilde{\tilde m}/\delta\tilde c$,
$\delta\bar p=\delta\tilde{\tilde p}/\delta\tilde c$, and all
functions depend on the scaling variable $z$ only. The least
stable solution corresponds to the smallest $\lambda$.
Integrating (numerically) the above equations we found
$\lambda\approx 5.41$, a fairly large value, indicating that not
only the scaling solution is locally stable in the limit $x\to 0$,
but also that it serves as a good approximation at finite $x$. The
correction to scaling functions $\delta\bar m(z)$ and $\delta\bar
p(z)$ are shown in Fig.~(\ref{dmdpfig}).

\emph{Numerical solution of the RSB equations.} To complete the proof
we have to demonstrate that the exact solution of Parisi equations
indeed flows to the fixed point, represented by our scaling
solution, which we had shown to be attractive. A simple way to do
this is to solve the exact RSB equations
[Eqs.~(\ref{diffeqmt},\ref{diffeqpt})] numerically on the interval
$x\in[x_0,1]$, using the scaling solution $\tilde p$ as the
initial conditions at the point $x_0\to 0$. If the function $c(x)$
computed in that way is consistent with our scaling ansatz, that
is $c(x)$ is approaching, for small $x$, the constant
$c=0.410802...$ (and $\tilde m$ is approaching its scaling
counterpart), then one can claim that the scaling region does
exist. A similar self-consistency check could be done on the
interval $x\in[0,x_0]$, though it is not as simple. In addition to
proving our point, the outlined procedure will allow us to compute
certain low energy quantities characterized by a finite time scale
$x=1$.

\begin{figure}[ptb]
\vspace{.05cm}
\includegraphics[width=3.2in]{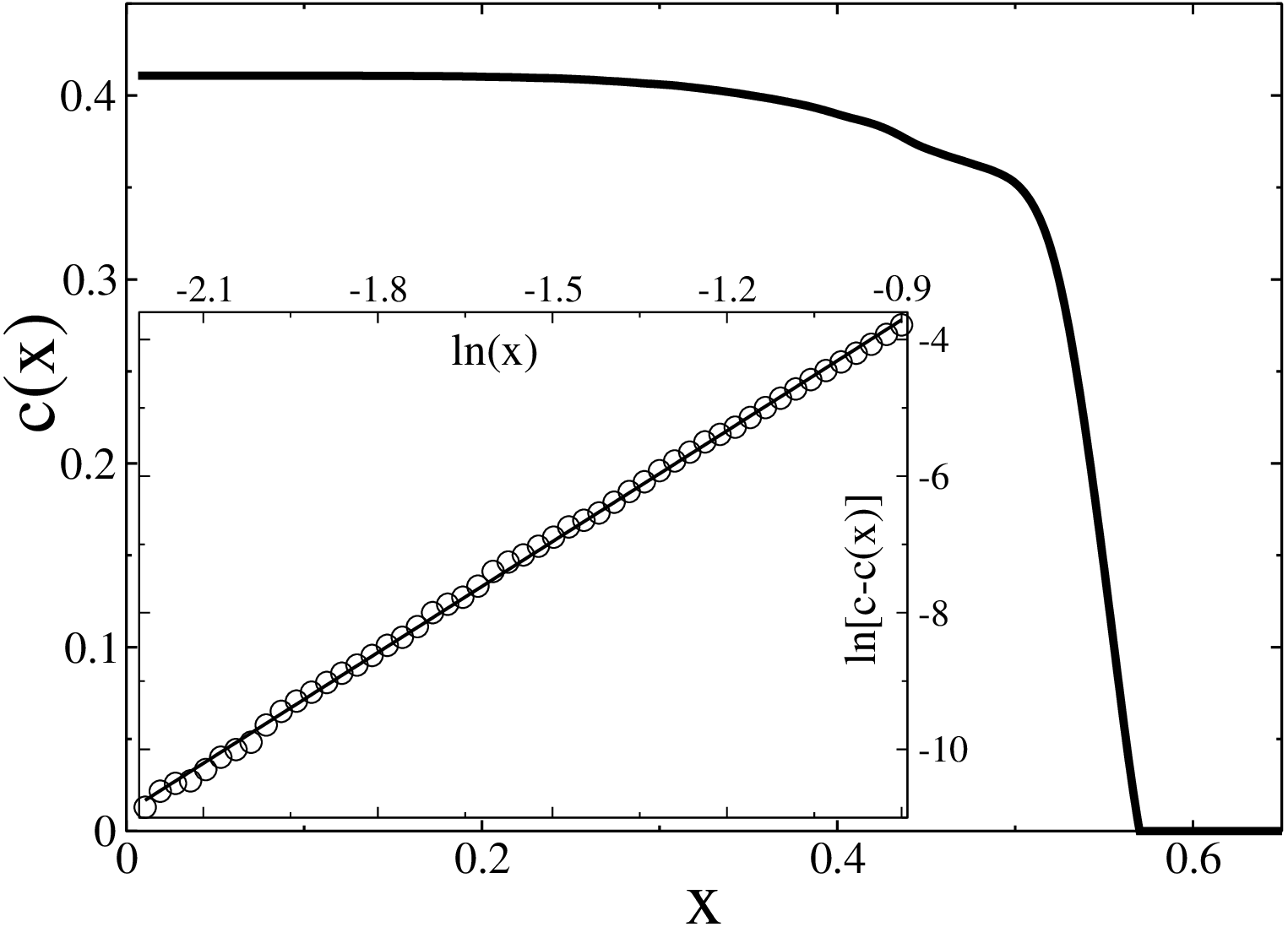} 
\vspace{.1cm}
\caption{Function $c(x)$ obtained by solving the Parisi equations 
numerically at $T=0$. It quickly approaches the constant value $c=0.410802...$ 
as $x\to0$. The inset shows $\ln{[c-c(x)]}$ plotted vs $\ln{x}$ (circles) 
along with the linear analysis prediction $5.41 \ln{x} +{\rm const}$ 
(solid line). Notice that the function $c-c(x)$ spans more than three 
orders of magnitude.}
\label{cxfig}
\end{figure}

In practice, when integrating 
Eqs.~(\ref{diffeqmt},\ref{diffeqpt}) numerically, it is convenient
to switch to a new variable $t=\ln{x}$ and avoid numerical
differentiation by computing $c(x)$ in successive iterations as
$c_{i+1}(x)=c_{i}(x)\int dz \tilde p(x,z)(\tilde m'(x,z))^2$. An
initial guess for $c(x)$ can be a rather arbitrary positive
function. The numerical solution quickly converges at small $x$,
while converging logarithmically slow for $x$ above and near the
breaking point $x_{max}$ \cite{crisanti02}. The process can be
speeded up by choosing judicially \cite{crisanti04} the initial guess
of $c(x)$, taking in consideration that it vanishes for $x > x_{max}$. 
The results are shown in Fig.~(\ref{cxfig}). The numerical solution 
reproduces the exact value of $c$ to five digits. In the inset we 
produce a log-log plot of $c-c(x)$ together with our stability 
analysis prediction of ${\rm const}\cdot x^{5.41}$. These results 
demonstrate the excellent approximating quality, as well as the 
validity, of the scaling ansatz and its correction.
\vspace{.2cm}

\begin{figure}[ptb]
\includegraphics[width=3.1in]{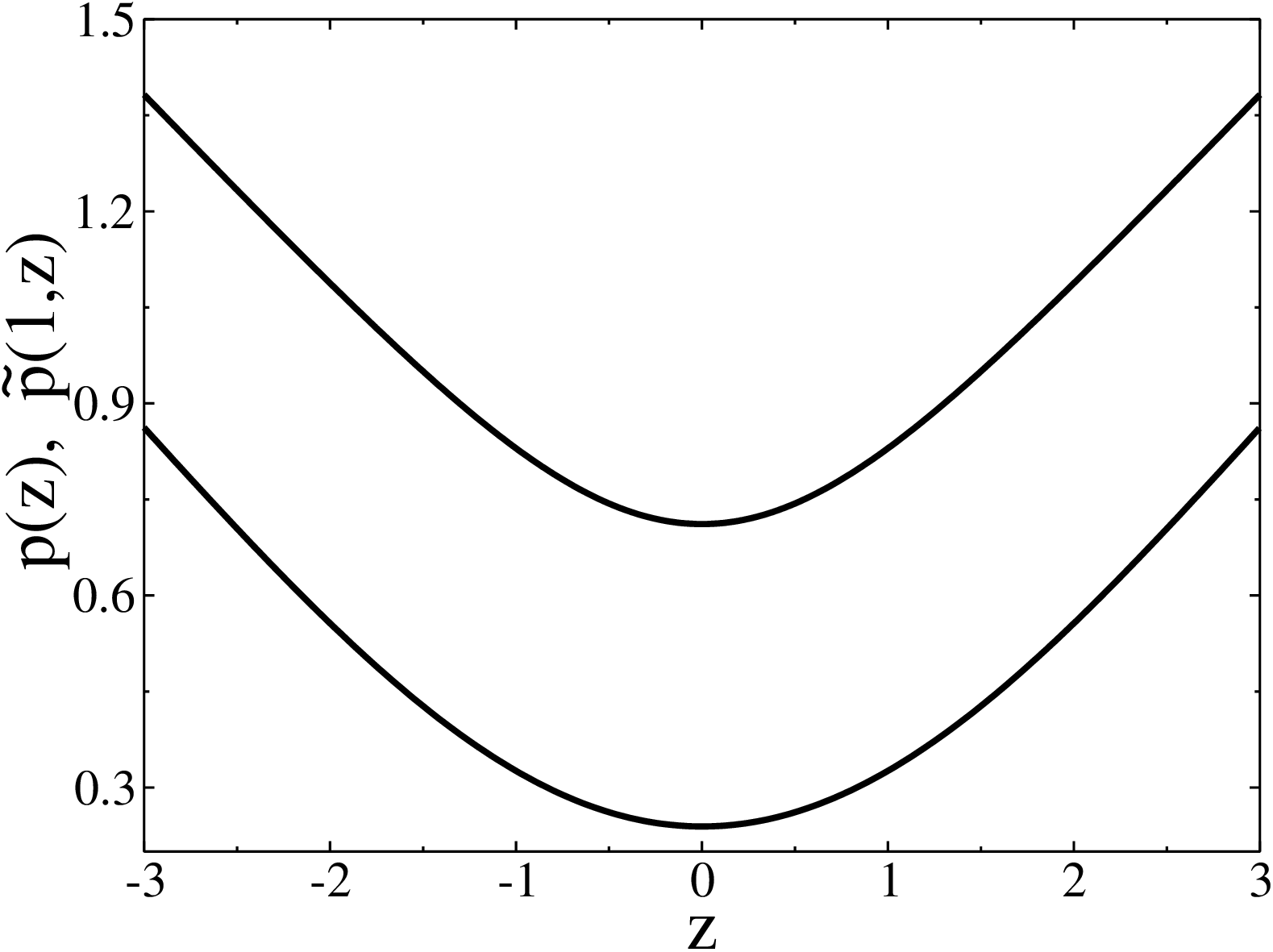} 
\vspace{.1cm}
\caption{Universal distribution functions 
$p(z)$ (instantaneous fields, bottom graph) and 
$\tilde p(1,z)$ (frozen fields, top graph) obtained by 
numerically solving the Parisi equations at $T=0$.
The instantaneous fields are enhanced, relative to the frozen fields,
by the value of the Onsager term $\beta\chi_0$ at large $z$.}
\label{pdistfig}
\end{figure}

Using the numerical solution of the Parisi equations, namely the
functions $\tilde p(1,z)$ and $\tilde m(1,z)$, we compute the
leading low temperature behavior of the zero field cooled 
susceptibility $\chi_0=\beta[1-q(1)]$
\begin{equation}
\beta\chi_0=\int dz \tilde p(1,z)[1-\tilde m^2(1,z)], \label{chi}
\end{equation}
and entropy $S$:
\begin{multline}
\beta^2 S=-(\beta\chi_0)^2/4\\
+\int dz \tilde p(1,z)\left[\ln(2\cosh{z})-z\tanh{z}\right].
\label{entropy}
\end{multline}
We found $\beta\chi_0=1.592\pm 0.002$ and $\beta^2 S=0.7210\pm
0.0002$, in agreement with known estimates \cite{crisanti02}, and,
to our best knowledge, with precision exceeding all other
methods. These values were computed for demonstrational
purposes only, and with some effort a far better accuracy can be
achieved. We also estimated the breaking point value to be
$x_{max}=0.56\pm 0.02$ which is in agreement with the literature.
Finally we compute a universal distribution function of the
instantaneous fields $p(z)=P(zT)/T|_{T\to0}$, where $P(y)$ is
experimentally measurable distribution of instantaneous
fields \cite{thomsen86}, corresponding to the tunneling density
of states in the context of charge glasses:
\begin{multline}
p(z)=\int \frac{dz'}{\sqrt{2\pi\beta\chi_0}} \tilde p(1,z')\\
\times\frac{\cosh{z}}{\cosh{z'}}
\exp{\left[-\frac{(z-z')^2}{2\beta\chi_0}
-\frac{\beta\chi_0}{2}\right]}.
\label{instfielddist}
\end{multline}
The distribution function $p(z)$ is plotted in 
Fig.~(\ref{pdistfig}) together with $\tilde p(1,z)$.
At large $z$ the distribution of instantaneous fields is shifted,
relative to the distribution of frozen fields due to effect of the 
Onsager reaction term \cite{thouless77,mueller04} 
by $\gamma\beta\chi_0\approx 0.479$. 

\emph{Conclusion.} To summarize, in this paper we presented a
scaling ansatz for the zero temperature solution of the SK model.
We assessed the stability of the scaling solution, both
analytically and numerically, and found that it becomes exact in
the limit of zero temperature and low energy, for $T\ll x\to0$ 
and $y\ll 1$. Our ansatz enabled us to compute the slope of 
the gap in the local fields distribution function numerically 
exactly, that is up to an arbitrary requested precision. Using 
the scaling solution as the initial conditions for the RSB 
equations at $x\to0$, we could compute, with high precision, 
the asymptotic behavior of finite time scale quantities, such 
as the entropy and zero field cooled susceptibility. 

We would like to note that the idea of
existence of a scaling solution of some sort for the SK model is
not new, one of the recent attempts was reported in Ref
 \cite{oppermann05}. In our paper however, for the first time, a
proposed scaling ansatz is proven to become asymptotically exact
in a certain limit. The presented method is not limited to the SK 
model only. It is expected to be applicable and very useful for a
wide variety of spin glass models which admit full RSB scenario,
such as the Ising $p$-spin model or a recently introduced mean field
description of the Coulomb 
glass \cite{mueller04,pankov05,mueller&pankov}.
The scaling ideas may also be promising for understanding quantum
glasses, which are thought to be relevant to the most challenging 
problems of strongly correlated systems \cite{dagotto05}.

\begin{acknowledgments}
The author acknowledges extremely enlightening discussions
with Vladimir Dobrosavljevic and Markus Mueller.
\end{acknowledgments}

\end{document}